\documentclass[twocolumn,pre,aps,showpacs,floatfix]{revtex4}

\usepackage[dvips]{graphicx}
\usepackage{amsmath}
\usepackage{amssymb}

\begin{document}

\title{L\'evy--Brownian motion on finite intervals: Mean first passage time analysis.}

\author{B. Dybiec}
\email{bartek@th.if.uj.edu.pl}
\affiliation{Marian Smoluchowski Institute of Physics and Mark Kac
Center for Complex Systems Research, Jagellonian University,
ul. Reymonta 4, 30--059 Krak\'ow, Poland}

\author{E. Gudowska-Nowak}
\email{gudowska@th.if.uj.edu.pl}
\affiliation{Marian Smoluchowski Institute of Physics and Mark Kac
Center for Complex Systems Research, Jagellonian University,
ul. Reymonta 4, 30--059 Krak\'ow, Poland}

\author{P. H\"anggi}
\email{peter.hanggi@physik.uni-augsburg.de}
\affiliation{Institute of Physics, University of Augsburg,
Universit\"atsstrasse 1, 86135 Augsburg, Germany}

\date{\today}

\begin{abstract}

We present the analysis of the first passage time problem on a finite interval for the
generalized Wiener process that is driven by L\'evy stable noises.
The complexity of the first
passage time statistics (mean first passage time, cumulative first passage time distribution)
is elucidated together with a discussion of the
proper setup of corresponding boundary conditions that correctly yield
the statistics of first passages for these non-Gaussian
noises. The validity of the method is tested numerically and compared against
analytical formulae when the stability index $\alpha$ approaches $2$, recovering in this limit
the standard results for the Fokker-Planck dynamics driven
by Gaussian white noise.

\end{abstract}

\pacs{02.50.Ey, 05.10.Gg, 05.10.Ln}

\maketitle

\section{Introduction}
Stochastic L\'evy processes serve as paradigms for the description
of many unusual transport processes leading to anomalous diffusion as
characterized by an anomalous mean squared
displacement, i.e.,
\begin{equation}
\langle(x(t)-\langle x(t)\rangle)^2\rangle=\langle(\Delta x)^2 \rangle \propto t^{\nu}
\label{di}
\end{equation}
which deviates with $\nu\neq 1$ from the linear dependence $\langle(\Delta x)^2\rangle\propto t$
that characterizes
normal diffusion. In the above
formula $\nu$ stands for the anomalous diffusion exponent that specifies the process
at hand as either behaving
 {\it subdiffusive}
(with $0<\nu<1$), {\it superdiffusive }($1<\nu$) or {\it ballistic} (for $\nu=2$)
\cite{ZASLAVSKY,Hughes,MetzlerPR,Drysdale}.
Among the class of L\'evy processes, the free L\'evy flights (LF) represent a special
class of discontinuous {\it Markovian} processes, for which the mean squared
displacement, as defined in equation (\ref{di})
diverges due to the heavy-tail distribution of the {\it independent increments} $\Delta x$.
In this case, the mean squared displacement is always superdiffusive; i.e.
$\langle(x(t)-\langle x(t)\rangle)^2\rangle\propto t^{2/\alpha}$, where $0<\alpha<2$ denotes the stability index
of the L\'evy--Brownian motion process, see below in Sect. II.
In contrast to the spatio-temporal coupling characterizing
general forms of non-Markovian, or more precisely, semi-Markovian \cite{Hughes}
L\'evy walks (LW) \cite{ZASLAVSKY,MetzlerPR,Drysdale}, L\'evy flights correspond
to the class of Markov processes
that emerge from a Langevin equation with
$\delta$-correlated, white L\'evy noise. Because LFs typically possess
a broad jump length distribution with an
asymptotic power law behavior their trajectories display
at all scales self-similar clustering
of local sojourns that become interrupted by long
jumps into the other location in the phase space where a new clustering forms.
Early discoveries of LF-like phenomena were related to intermittent chaotic systems and description
of the motion of the fluid particles in fully developed turbulence \cite{ZASLAVSKY,Hughes,MetzlerPR}.
Nowadays, their applications range from description of the dynamics in plasmas,
diffusion in the energy space, self-diffusion in micelle systems and transport
in polymer systems under conformational motion \cite{ZASLAVSKY,MAI} to the
spectral analysis of paleoclimatic \cite{DIT} or economic data \cite{SAN}.
Despite the ubiquitous use of LFs as phenomenological models for noise sources,
their influence on the kinetics subjected to boundary conditions has been
addressed scarcely only. For free normal diffusion, the knowledge of the Green function,
together with the local boundary conditions, is sufficient to determine the
first passage time statistics. The same information can be also obtained
by use of the method of images or by solving the corresponding, local boundary
value problem of the diffusive Fokker-Planck equation
\cite{cox1965,goelrichter1974}.

In particular, for processes
driven by white Gaussian noise these boundary conditions for {\it reflection} or {\it absorption}
are locally defined and well known \cite{goelrichter1974,HTB90}. However,
for the case of LFs, the method of images fails, yielding
results that contradict the Sparre-Andersen theorem \cite{chechkin,Metzler2004}.
By virtue of the latter \cite{SPARRE}, for any discrete time random walk starting out
at $x_0\not=0$ with the step length sampled from a continuous, symmetric distribution,
the first passage time density decays
 asymptotically as $t^{-3/2}$. In order to further explore the intricate problem of proper
 boundary conditions for absorption and reflection for non-Gaussian white noise composed of
 independent L\'evy flight increments, we elucidate with this work the situation of a free,
 overdamped L\'evy--Brownian motion, being restricted to a bounded domain of attraction.

\section{Restricted L\'evy--Brownian motion}
We consider the free Brownian motion on a restricted, finite interval that is driven by L\'evy stable noise.
More specifically, the dynamical evolution of a stochastic state variable $x(t)$ is described in terms of the
Langevin equation
\begin{equation}
\frac{dx(t)}{dt}=\zeta(t)\;,
\label{lang}
\end{equation}
where $\zeta(t)$ denotes a L\'evy stable white noise process which is
composed of {\it independent} differential
increments that are distributed according to the stable density with the index
$\alpha$; i.e. $L_{\alpha,\beta}(\zeta;\sigma,\mu=0)$.
Put differently, $\zeta(t)$  stands for the generalized white noise process which is obtained from
the time derivative of the corresponding L\'evy--Brownian (Markovian)
stable process. The parameter choice $\alpha=2$ yields the usual, $\delta$-correlated Gaussian white noise.
In contrast, with $ 0< \alpha < 2$
the corresponding L\'evy white noise is generated from  L\'evy--Brownian motion  possessing
discontinuous sample paths with infinite variance and the higher cumulants. Its statitiscal properties can be however characterized by fractional moments of order $\nu$ which exist and are finte for $\nu<\alpha<2$ \cite{west}.
Because it is composed of independent increments,
this L\'evy stable white noise also constitutes a singular white
noise process whose autocorrelation again is formally $\delta$-correlated  \cite{west}.
Here, the
parameter $\alpha$ denotes the stability index, yielding the
asymptotic power law for the jump length distribution being
proportional to $|\zeta|^{-1-\alpha}$. The parameter $\sigma$
characterizes a scale, $\beta$ defines an asymmetry (skewness) of
the distribution, whereas $\mu$ denotes the location parameter.
We deal only with strictly stable distributions not exhibiting
a drift regime; this implies a vanishing location parameter $\mu=0$
throughout the remaining part of this work. For $\alpha\not=1$, the
characteristic function $\phi(k) = \int_{-\infty}^\infty
e^{-ik\zeta} L_{\alpha,\beta}(\zeta;\sigma,\mu=0) d\zeta$ of an
$\alpha$-stable random variable $\zeta$ can be represented by
\begin{equation}
\phi(k) = \exp\left[ -\sigma^\alpha|k|^\alpha\left( 1-i\beta\mbox{sign}(k)
\tan\frac{\pi\alpha}{2} \right)\right],
\label{charakt}
\end{equation}
while for $\alpha=1$ this expression reads
\begin{equation}
\phi(k) = \exp\left[ -\sigma|k|\left( 1+i\beta\frac{2}{\pi}\mbox{sign} (k) \ln|k| \right) \right] \;.
\label{charakt1}
\end{equation}
The three remaining parameters vary within the regimes
$
\alpha\in(0,2],\;
\beta\in[-1,1],\;
\sigma\in(0,\infty)
$.

The stochastic differential equation in (\ref{lang}) yields normal, free Brownian motion when
$\alpha=2$, and free super-diffusion when $\alpha\in(0,2)$. The numerical integration of
equation~(\ref{lang}) has been performed by use of standard techniques of integration
of stochastic differential equation
with respect to the L\'evy stable measures and studied by use of the
Monte Carlo methods~\cite{newman1999}. In particular, the error bars visible in the figures
were calculated using the bootstrap method. The position of a L\'evy--Brownian particle has been
obtained by a direct integration of Eq.~(\ref{lang}) leading to the following approximation
\cite{weron1995,weron1996,ditlevsen1999,janicki1994,janicki1996,dybiec2004}
\begin{equation}
x(t) = \int_{0}^{t} \zeta(s)ds \approx
\sum\limits_{i=0}^{N-1}(\Delta s)^{1/\alpha}\zeta_i,
\label{lcalka}
\end{equation}
where $\zeta_i$ are independent random variables distributed with the probability density function (PDF)
$L_{\alpha,\beta}(\zeta;\sigma,\mu=0)$ and $N\Delta s=t$. All our
illustrations in this work are based on trajectory calculations
sampled from the Langevin equation (\ref{lang}). Absorbing boundary conditions
have been realized by stopping the trajectory whenever it reached the boundary, or, more typically, it has
jumped beyond that boundary location. The condition of reflection has been assured by wrapping
the hitting (or crossing) trajectory around the boundary location, while preserving its assigned length.
The Appendix A provides some further details on the numerical
scheme for stochastic differential equations driven by L\'evy white noise.
We also like to emphasize that our results omit cases when
$\alpha= 1$ with $\beta\ne0$. In fact, this parameter set is known to induce instabilities in
the numerical evaluation of corresponding trajectories
\cite{weron1995,weron1996,ditlevsen1999,janicki1994,janicki1996,dybiec2004}.
Thus, for our numerics with $\beta\ne0$,
the parameter value $\alpha =1$ has been excluded from the consideration.

\section{Mean first passage times for $\alpha$-stable noises}

\begin{figure}[!ht]
\includegraphics[angle=0, width=6.0cm]{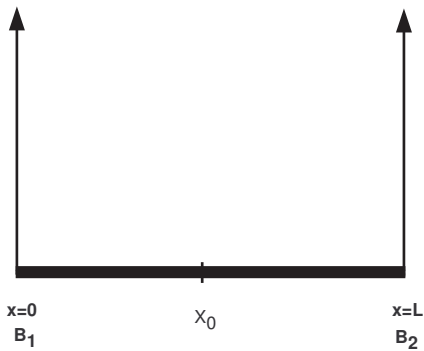}
\caption{Setup for the investigation of the first passage time analysis
of L\'evy white noise driven free Brownian motion being confined between the
two boundaries located at $B_1$ and $B_2$, respectively.}
\label{setup}
\end{figure}

Our main objective is the investigation of the mean first passage time (MFPT) of
L\'evy--Brownian motion on a finite interval as illustrated with Fig.~\ref{setup}.
The boundaries at both ends will be assumed to be of either {\it absorbing} or also of the {\it reflecting} type.
With generally non-Gaussian white noise the knowledge of the boundary location alone cannot
specify in full the corresponding boundary conditions for absorption or reflection, respectively.
In particular, the trajectories driven by non-Gaussian white noise depict discontinuous jumps, cf.
Figs.~\ref{normal_tr} and~\ref{flight_tr} below. As a consequence, the location of the boundary itself is not hit by
the majority of discontinuous sample trajectories. This implies that regimes beyond the location of the boundaries
must be properly accounted for when setting up the boundary conditions. Most importantly, returns
(i.e. so termed re-crossings of the boundary location) from excursions
beyond the specified state space back into this very finite interval
where the process proceeds must be
excluded. Thus, the problem of proper formulation of boundary conditions in such cases poses an open challenge
that has not been addressed with sufficient care in the prior literature
\cite{gitterman2000,gitterman2004}. In contrast to the case with normal diffusion (i.e. when $\alpha=2, \beta =0$),
these boundary conditions are of a {\it nonlocal} nature; as a consequence, an analytical
investigation of the mean first passage problem
becomes very demanding and cumbersome. In this work, we therefore restrict ourselves
predominantly (for $\alpha < 2$) to detailed, precise numerical simulations.

The setup for the studies is schematically depicted in Fig.~\ref{setup}.
The stochastic motion of a free L\'evy particle is confined in an interval
specified by the two boundaries $B_1,B_2$. The dynamics of the
LFs trajectories derived from Eq.~(\ref{lang}) are consequently confined to this state space
in-between $B_1$ and $B_2$.

\begin{figure}[!ht]
\includegraphics[angle=0, height=6.0cm]{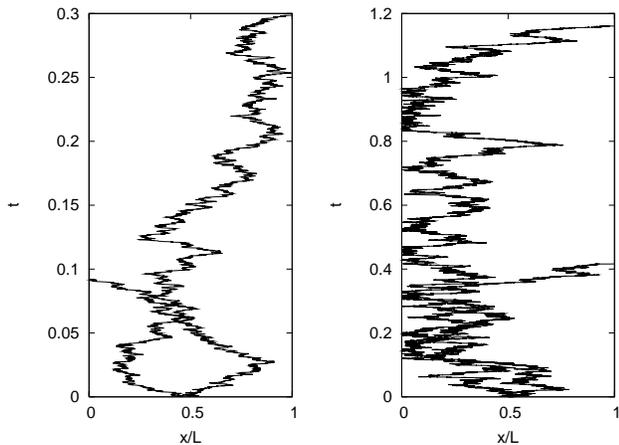}
\caption{Left panel: Typical sample trajectories of confined, normal diffusion, i.e. $\alpha=2$, with
two absorbing AA boundaries. Right panel: Confined normal Brownian motion between a left-sided
reflecting boundary (R) and a right-sided absorbing boundary (A), denoted as RA in the text.
The particles start out at midpoint
$x/L=0.5$, undergoing state-continuous stochastic motion with evolving time $t$.}
\label{normal_tr}
\end{figure}

\subsection{The test case: normal diffusion}
\begin{figure}[!ht]
\includegraphics[angle=0, width=8.0cm]{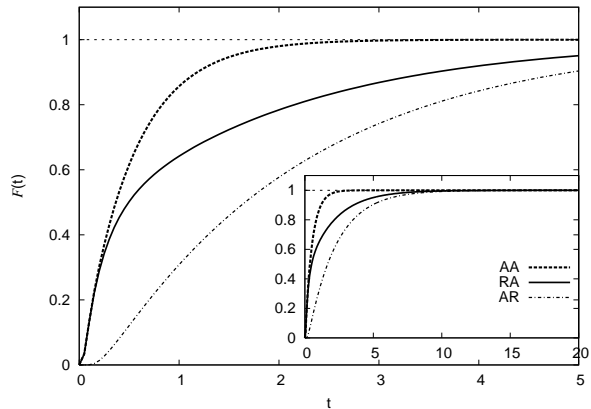}
\caption{First passage time statistics for free normal diffusion ($\alpha=2$) occurring
between two boundaries located at $x=0$ and $x=5$.
The cumulative first passage time distribution $\mathcal{F}(t)$ is depicted
versus the first passage time variable $t$. The initial motion starts out at
$x_0=3.5$ and the noise strength is $\sigma^2=5$.
The chosen time step is $\Delta t=10^{-5}$ and the results have been averaged over $N=3\times 10^4$
realizations. The
simulation results are presented
for AA, RA and AR configurations (from the top to the bottom) along
with lines representing the analytical formulae for these cases. The numerical data match perfectly
the analytical results, plotted on superimposed lines. The inset depicts this agreement on an extended time span
up to time $t=20$.}
\label{cdfgaus}
\end{figure}

For $\alpha=2$ (see Fig.~\ref{normal_tr}), white L\'evy stable noise becomes equivalent to Gaussian
white noise and the corresponding Langevin eqution (\ref{lang}) describes a free Brownian motion (Wiener process) for which the probability density functions (PDFs) of the first
passage times distributions are known explicitly from the
literature \cite{goelrichter1974,cox1965}. Here, due to reasons of convenience,
instead of estimating the PDFs itself we simulate the equivalent cumulative distribution functions (CDFs)
of first passage times. For the sake of simplicity
it has been assumed that the left boundary is located at $x=0$ and the other one at
 $x=L$. In the following we typically use, although not exclusively, for the initial condition, $x_0$,
 the center of the interval, $x_0=L/2$.
In the case when both $B_1$ and $B_2$ are absorbing boundaries, denoted as AA,
the cumulative
distribution function $\mathcal{F}$ of first passage times has been
obtained by integration of the first passage time density $f(t)$, i.e.
 $\mathcal{F}(t)=\int_0^tf(t')dt'$ \cite{cox1965,goelrichter1974,HTB90}:
\begin{equation}
\mathcal{F}(t)= 1-\frac{2}{\pi}\sum\limits_{j=1}^{\infty}\frac{1-\cos j\pi}{j}
\sin\frac{j\pi x_0}{L}\exp\left[ -\left( \frac{j\pi\sigma}{L} \right)^2 t \right] \;.
\label{aacdfeq}
\end{equation}
Analogously, for a left absorbing boundary (A) and a reflecting right boundary (R), denoted as AR,
 the corresponding CDF of first passage times reads \cite{cox1965,goelrichter1974,HTB90}
\begin{eqnarray}
\mathcal{F}(t) & = & \frac{4}{\pi}\sum\limits_{j=0}^{\infty}\frac{1}{2j+1}\sin\frac{(2j+1)\pi}
{2}\cos\frac{(2j+1)\pi(L-x_0)}{2L} \nonumber \\
& \times & \left[ 1 - \exp\left[ -\left( \frac{(2j+1)\sigma\pi}{2L} \right)^2 t \right] \right] \;.
\label{arcdfeq}
\end{eqnarray}
Finally, for a left reflecting boundary (R) and a absorbing right boundary (A), denoted as RA,
 the corresponding CDF of first passage times reads \cite{cox1965,goelrichter1974,HTB90}
\begin{eqnarray}
\mathcal{F}(t) & = & \frac{4}{\pi}\sum\limits_{j=0}^{\infty}\frac{1}{2j+1}\sin\frac{(2j+1)\pi}
{2}\cos\frac{(2j+1)\pi x_0}{2L} \nonumber \\
& \times & \left[ 1 - \exp\left[ -\left( \frac{(2j+1)\sigma\pi}{2L} \right)^2 t \right] \right] \;.
\label{racdfeq}
\end{eqnarray}
where the scaling parameter $\sigma$ stands for the amplitude of the noise
intensity for the additive white Gaussian
noise; i.e. $\langle \zeta(t) \zeta (s) \rangle = 2\sigma^2 \delta (t-s)$.
Note, that for a particle starting its motion at $x_0=L/2$,
$\mathcal{F}(t)$ for a RA boundary setup is the same as for the symmetrically chosen
 AR situation, cf. the formulae above.
To test our employed software, we evaluated numerically the
first passage times distributions with the same boundary conditions
 as specified above. Eq.~(\ref{lang}) has been integrated numerically for
$B_1=0$, $B_2=5$, $x_0=3.5$ with $\Delta t=10^{-5}$ and averaged over
$N=3\times10^4$ realizations. Figure ~\ref{cdfgaus} depicts the numerical results along
with their corresponding analytical expressions for the corresponding
compound distribution functions (CDFs) of first passage times
 for a L\'evy-Gaussian case ($\alpha=2$). We find a
perfect agreement between the theoretical and the numerical results.
On purpose, we have chosen here an asymmetric starting point in order to impose an explicit
difference between the AR and the corresponding RA setup, respectively.

\subsection{Confined L\'evy--Brownian motion }

After having tested the numerical algorithm, we next study the mean first passage time for confined
L\'evy--Brownian motion on a finite interval.

\subsubsection*{MFPT for symmetric L\'evy noise}
\begin{figure}[!ht]
\includegraphics[angle=0, height=6.0cm]{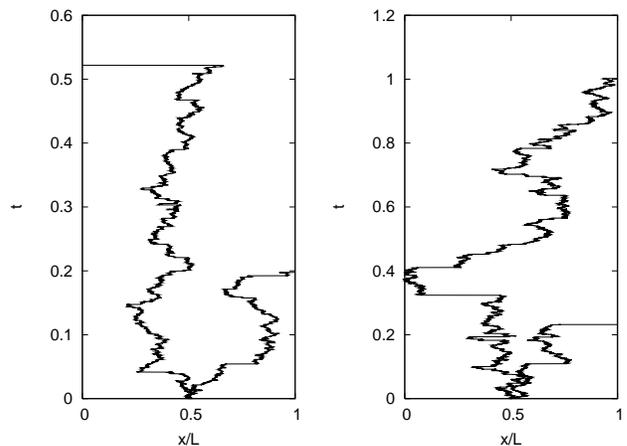}
\caption{Sample trajectories for AA (left panel) and RA (right panel) with the two
boundaries located at $B_1=0$ and $B_2=L$ for symmetric L\'evy flights with $\alpha=1.4,\beta = 0 $.
For $\alpha<2$ large excursions of the trajectory are more
probable than for the normal Brownian motion. A test ~``particle'' can skip over the border,
i.e. it can escape from the domain of motion with a single jump.}
\label{flight_tr}
\end{figure}

Using the discussed simulation procedure we start out with the case of confined L\'evy--Brownian motion
with {\it symmetric} stable L\'evy noise, i.e. we set $\beta=0$.
Exemplary trajectories are depicted in Fig.~\ref{flight_tr}.
for the stability index $\alpha= 1.4$. In clear contrast to normal Brownian motion we detect
discontinuous jumps, characterizing random jumps over large distances (note the horizontal
excursions in Fig.~\ref{flight_tr}). As a consequence, the boundaries
at $B_1,B_2$ are typically not hit, but rather crossed in a flight-like manner, being characteristic for L\'evy
distributed jumps. In fact, for $\alpha<2$ large excursions of the trajectory are more probable
than for normal Brownian motion. In effect, a test ``particle'' can skip over the border,
i.e. it can escape from the domain of motion via a single jump.

As discussed in the literature \cite{ZASLAVSKY,MetzlerPR,chechkin},
the scaling nature of the jump length PDFs causes a clustering
of L\'evy flights. Random localized motion is occasionally interrupted by
long sojourns on all length scales and, additionally, there are clusters of
local motion within clusters. Anomalous trajectories of L\'evy
flights with stability index $\alpha < 2$ influence also the
boundary condition for the mean first passage time and its corresponding statistics.

From the perspective of a random walk approximation to the Langevin
equation in (\ref{lang}), the behavior of a L\'evy walker is drastically different for
L\'evy jump length statistics as compared with a traditional
Gaussian case: The increments of the normal (state-continuous) diffusion process $x(t)$ are
characterized by statistics which excludes with very large probability the
occurrence of long jumps. Therefore, the walker is more
probable to approach -- and eventually hit -- a point-like boundary (cf. Fig.~\ref{normal_tr}).
In contrast (see Fig.~\ref{flight_tr}),
with the L\'evy jump statistics ($\alpha< 2$), a meandering
particle may easily cross the local boundary during its long jump
and may recross into the finite interval many times, unless the particle is
immediately absorbed upon crossing for the first time the boundary $B_1$ or $B_2$, respectively.
This brings about a formulation of the boundary condition that necessarily must be non-local in nature.

Indeed, the first passage time problem for Eq.~(\ref{lang}) can also be rephrased in
terms of the fractional Fokker-Planck
equation~\cite{chechkin,zaslavsky94,fogedby98,yanovsky2000,elizar03,dubkov05}.
For a free L\'evy--Brownian motion this equation with $\beta=0$ assumes the form:
\begin{eqnarray}
\frac{\partial p(x,t)}{\partial t}=
\sigma^\alpha\frac{\partial^{\alpha}}{\partial|x|^{\alpha}}p(x,t) \;.
\end{eqnarray}
The boundary conditions for the first passage time problem
associated with two absorbing boundaries at $B_1$ and $B_2$ are now nonlocal; reading,
\begin{eqnarray}
p(x,t)=0 \;\mbox{for}\; x\leqslant B_1 \; \;\mbox{and}\; \; p(x,t)=0\;\mbox{for}\;x\geqslant B_2.
\end{eqnarray}
Due to discontinuous character of trajectories of L\'evy processes we note that
the usual form of boundary conditions, i.e. $p(x=0,t)=0$ and $p(x=L,t)=0$, incorrectly employed
in the literature \cite{gitterman2000,buld01,buldpre01,gitterman2004}
is expected to lead to erroneous results; such a boundary condition
does not account for the fact that the process can skip the location of the boundary without hitting it.
The corresponding MFPT can be numerically integrated, yielding
\begin{eqnarray}
MFPT=\int^{\infty}_0-t\;dt\int^{B_2}_{B_1}\dot{p}(x,t) dx,
\end{eqnarray}
which after a partial integration equals
\begin{eqnarray}
MFPT=\int^{\infty}_0dt\int^{B_2}_{B_1}p(x,t) dx \;.
\end{eqnarray}

This expression has been tested versus the statistical definition based
on histogram analysis of the corresponding trajectories Eq.~(\ref{lang}), for which the
appropriate boundary conditions were imposed.

\begin{figure}[!ht]
\includegraphics[angle=0, width=8.5cm]{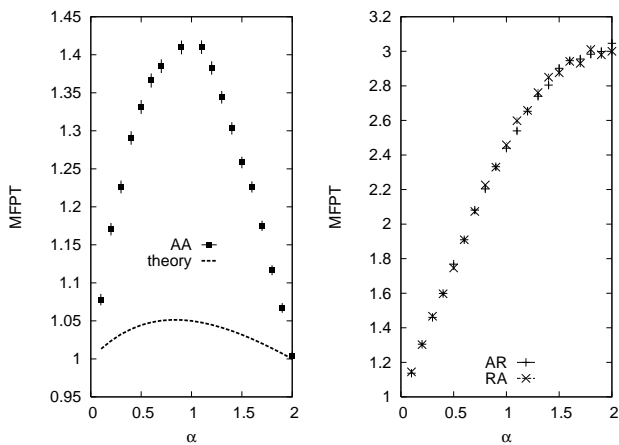}
\caption{Mean first passage time versus the stability index $\alpha$ of confined Brownian
motion driven by stable symmetric L\'evy noise. We depict the case of two absorbing boundaries,
AA (left panel), and the situation
of a reflecting and absorbing barrier, AR/RA (right panel).
The initial starting point has been chosen at midpoint $x_0=2$,
yielding identical results for AR and RA, respectively. The simulations parameters are:
$\beta=0$, $\sigma=\sqrt {2}$, time step $\Delta t=10^{-4}$, number of realizations $N=2\times10^2$.
The boundaries are located at
$B_1=0$ and $B_2=4$.
The theory-result from Ref. \cite{gitterman2000} (Eq.~(39) therein) is plotted as a dashed line.}
\label{mfpt_b00}
\end{figure}

Likewise, we also have studied the case where one of the two boundaries becomes reflecting.
The reflecting boundary is imposed on the trajectory simulations via an infinite high hard wall,
yielding immediate reflection \cite{bala88}. For the numerical implementation of this reflecting case
see below Eq.~\ref{lcalka}.
In Fig.~\ref{mfpt_b00}
we compare our numerical results for the case of two absorbing boundaries, case AA, and as well,
for the symmetric situation, AR = RA, of a reflecting boundary and an absorbing boundary with the initial
starting value chosen at midpoint.


Fig.~\ref{mfpt_b00} depicts the numerical results for the MFPT for symmetric
L\'evy stable noise after implementing numerically the appropriate boundary conditions as discussed
above. In the left panel of Fig.~\ref{mfpt_b00} the numerical results for the MFPT are
compared with the theoretical findings for the MFPT $\equiv T$ in this superdiffusive case, reading from
Eq.~(39) in Refs. \cite{gitterman2000,gittermanfootnote},


\begin{equation}
T=\frac{4}{\pi D_\alpha}\left( \frac{L}{\pi}
\right)^\alpha\sum\limits_{m=0}^{\infty}\frac{(-1)^m}{(2m+1)^{1+\alpha}},
\end{equation}
where $D_\alpha=\sigma^\alpha$.

The maximal MFPT is assumed for $\alpha \simeq 1$.
The behavior of the MFPT assumes a bell-shaped behavior around $\alpha=1$; it reflects the different interplay of
the probability of finding long jumps versus a decreasing stability index $\alpha <2$, implying a decreasing
noise intensity $\sigma^{\alpha}$. Put differently, the occurrence of
long jumps beyond the finite interval is dominating the escape over a decreasing noise intensity for
for $\alpha<1$.

The expected discrepancy between these analytical
results and numerical estimation is due to assumed local boundary condition of vanishing
probabilities $p(x=0)=p(x=L)=0$, which are correct only for normal Brownian motion; i.e. $\alpha=2$.
Indeed, our numerical result just coincides precisely at this very special value.
In our simulation of the Langevin equation in (\ref{lang}) the whole exterior of the prescribed
interval $(B_1=0,B_2=4)$ is absorbing throughout, while for the
analytical calculations the flawed, point-like boundary conditions are assumed
\cite{gitterman2000}.

Clearly, the extension of the absorbing regime to
the whole two semi-lines outside the confining interval yields the physically correct value
 that accounts for the escape of the particle from the
interval via long jumps. The numerical results for $\alpha < 2 $ systematically
 exceed the theory-result of Ref. \cite{gitterman2000}.

The right panel of Fig.~\ref{mfpt_b00} depicts the case for the MFPTs
with AR and RA boundaries, respectively. Due to symmetric chosen initial condition
and the symmetric stochastic driving the results for AR and RA become identical.

\subsubsection*{MFPT for asymmetric L\'evy noise}

\begin{figure}[!ht]
\includegraphics[angle=0, width=8.5cm]{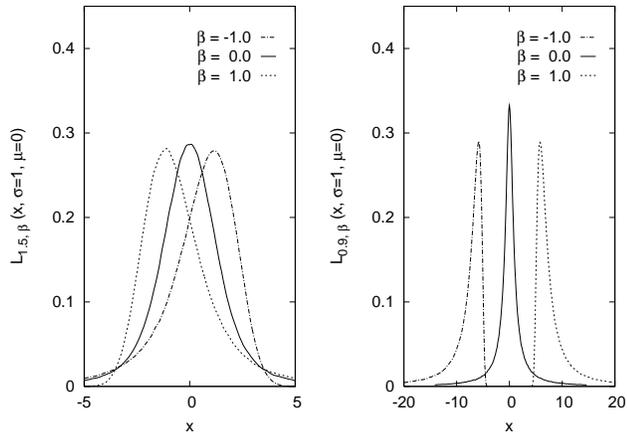}
\caption{ Probability density functions for the $\alpha$-stable variables $x$
with $\alpha = 1.5$ (left panel) and $\alpha = 0.9$ (right panel). The symmetric case is for $\beta = 0$,
while $\beta=\pm 1$ corresponds to asymmetric densities.
Note the differences in the positions of the maxima for $\alpha<1$ and $\alpha>1$.
Most importantly, the support of the densities for the fully asymmetric cases with $\beta=\pm 1$ and $\alpha < 1$
(right panel) is covering not the whole axis; it assumes only negative values for $\beta = -1$ and only positive
values for $\beta = 1$. This in turn causes the discontinuous behavior depicted with Fig.~\ref{mfpt_b10}. }
\label{density}

\end{figure}

\begin{figure}[!ht]
\includegraphics[angle=0, width=8.5cm]{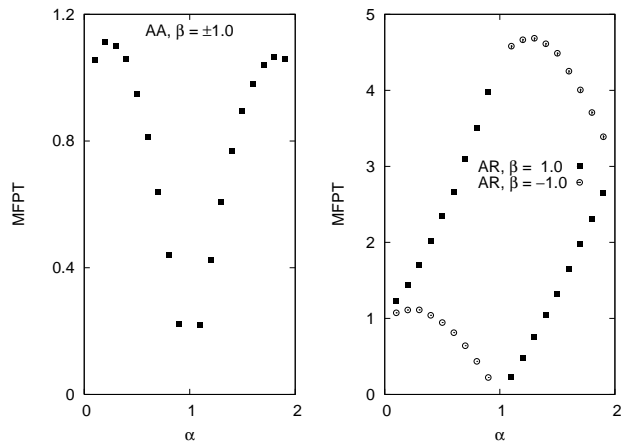}
\caption{MFPTs versus the stability index $\alpha$
for asymmetric L\'evy noise for different boundary conditions:
 AA (left panel), AR (right panel).
The simulations parameters are as in Fig.~\ref{mfpt_b00}. Note that the results
for the AR setting with $\beta=-1$ and $\alpha<1$ are the same as for the
AA settings with $\beta\pm1$ and $\alpha<1$.}
\label{mfpt_b10}
\end{figure}

Asymmetric L\'evy noise is characterized by a non-vanishing skewness parameter
$\beta \neq 0$ (see the exemplary probability density functions in Fig.~\ref{density}).
Fig.~\ref{mfpt_b10} presents the results of MFPT evaluations for fully asymmetric L\'evy stable
noise driving with $|\beta|=1$.
The left panel displays results for the L\'evy--Brownian motion with $\beta= 1$ within the AA boundaries setups.
Due to the imposed symmetry of a starting point, the results coincide with those
for the L\'evy noise driving with $\beta=-1$. The right panel depicts the
results for AR boundaries with $\beta=\pm 1$.
As can be inferred by inspection of Fig.~\ref{density}, due to the skewness character of the L\'evy stable distribution,
the results for AA boundaries with $\alpha<1$ and $\beta=1$ are the same as for AR boundaries with
$\alpha<1$ and $\beta=-1$. The effect is caused by a visible shift of the
probability mass apart from $x=0$ to the left (or to the right) for skewed distributions with $\alpha<1$.
In contrast, these results differ for $\alpha>1$ where both boundary setups, i.e. AA and AR,
lead to different values for MFPTs.


\begin{figure}[!ht]
\includegraphics[angle=0, width=8.5cm]{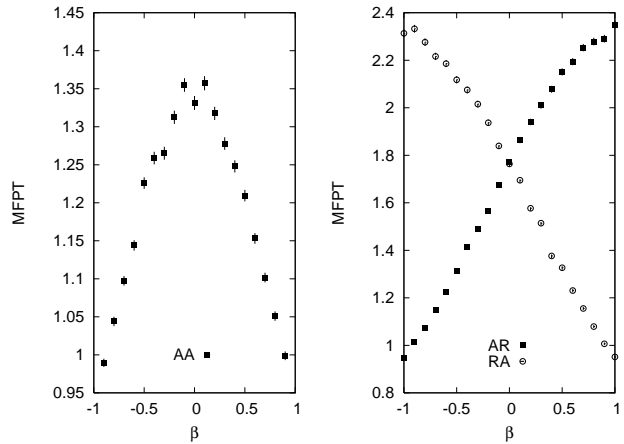}
\caption{MFPTs versus the skewness parameter $\beta$ for AA (left panel) and AR (right panel) boundaries
for L\'evy--Brownian motion with $\alpha=0.5$. The simulations parameters are like in Fig.~\ref{mfpt_b00}.}
\label{mfpt_a05}
\end{figure}

As can be intuitively expected, the cumulative first passage time distributions (CDFs)
of the first passage times as governed by
Eq.~(\ref{lang}) for the AA boundaries with a symmetric starting point, i.e.
$x_0=L/2$, are invariant under the transformation $\beta\to-\beta$. Therefore,
the quantities derived from the numerically determined CDFs are symmetric
around $\beta=0$, cf. the left panel of Fig.~\ref{mfpt_a05}. Furthermore, because of the
chosen symmetric initial conditions the cumulative first passage time distributions evaluated for the RA
boundaries setup can be constructed from the appropriate distribution for the
AR case by exchanging $\beta$ with $-\beta$. It is caused by the fact that the
asymmetry induced by a nonzero skewness parameter can be compensated by an exchange
of boundaries, cf. left panel of Fig.~\ref{mfpt_a05} and Fig.~\ref{density}.

\subsection{First passage time statistics}

\begin{figure}[!ht]
\includegraphics[angle=0, width=8.5cm]{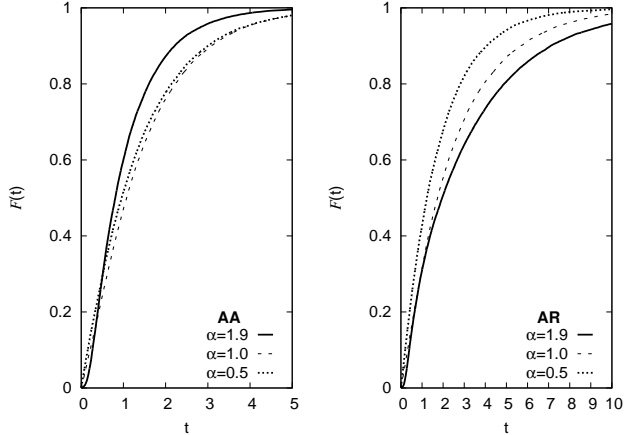}
\caption{Cumulative first passage time distribution function
for symmetric L\'evy--Brownian motion on a restricted interval, i.e. $\beta=0$, with
$\alpha=1.9,1.0,0.5$ for AA (left panel) and AR (right panel) boundaries.
The results have been simulated for $\sigma=\sqrt{2}$, $\Delta t=10^{-4}$ and have been
averaged over $N=2\times10^2$ realizations. The boundaries $B_1,\;B_2$ are
located at $x=0$ and $x=4$. The particle starts out at midpoint; i.e. $x_0=2$.}
\label{alpha}
\end{figure}

\begin{figure}[!ht]
\includegraphics[angle=0, width=8.5cm]{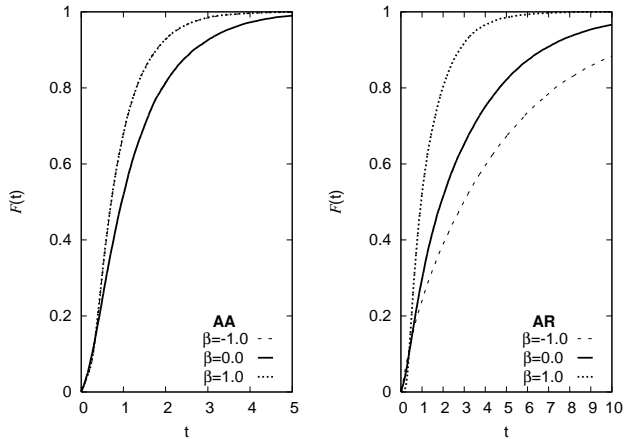}
\caption{Cumulative first passage time distribution function versus first passage time $t$
for L\'evy noises with $\alpha=1.5$ and $\beta=-1.0,0.0,1.0$
for AA (left panel) and AR (right panel) boundaries. The simulations details
are the same as in Fig.~\ref{mfpt_b00}. Note that the results for the AA settings with $\beta=\pm1$ overlap.}
\label{beta}
\end{figure}

The simulation data for the mean first passage time yield as well the results for the cumulative
first passage time distribution functions (CDFs). Typical such distributions are depicted in Fig.~\ref{alpha}
for symmetric $\alpha$-stable noise and in Fig.~\ref{beta} for asymmetric stable L\'evy white noises with a
stability index $\alpha=1.5$. With symmetric noise we observe that the escape, starting out at midpoint,
is speeded up with two absorbing boundaries AA as compared to the situation with AR=RA.
The reflecting boundary clearly slows
down the ultimate escape from the finite interval.

For escape driven by
asymmetric white L\'evy noise, see Fig.~\ref{beta}, we note that for the case of two absorbing boundaries AA
(left panel) the CDFs become identical for $\beta=-\beta$, given the midpoint starting value.
We also can detect a more rapid saturation with $\beta\neq 0$
as compared to the fully symmetric situation with $\beta=0$;
implying a somewhat faster escape scenario.
This fact originates from the skewness in the distribution of
jump values, implying a faster escape towards the corresponding absorbing boundary.

With an AR boundary setup the situation becomes more intricate.
Now, depending on the choice of the asymmetry parameter
the escape can be enhanced, reflecting the skewness of the
corresponding stable distribution
and the relative character of the boundary setup.

\begin{figure}[!ht]
\includegraphics[angle=0, width=8.5cm]{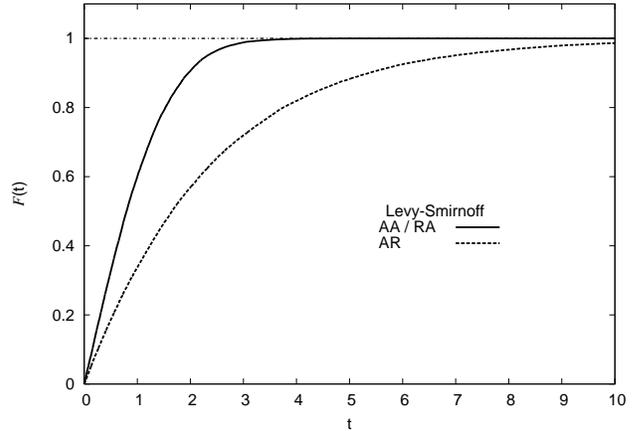}
\caption{The cumulative first passage time distribution $\mathcal{F}(t)$ versus first passage time $t$
for L\'evy--Smirnoff driven L\'evy--Brownian motion, i.e. $\alpha=0.5$ and
$\beta=1$ and a scale parameter $\sigma=\sqrt{2}$.
Because the L\'evy--Smirnoff statistics assumes only positive values the
distributions $\mathcal{F}(t)$ for AA and RA configurations become the same.
The simulations parameters are as in Fig.~\ref{mfpt_b00}.}
\label{smirnoff}
\end{figure}

This complexity is elucidated further in panel \ref{smirnoff}
for the case of stable L\'evy--Smirnoff noise where
$\alpha=0.5, \beta=1$. The CDF $\mathcal{F}(t)$ is depicted in Fig.~\ref{smirnoff}.
For $\alpha<1$ and $|\beta|=1$ L\'evy stable distributions are
totally skewed, taking on either only positive values
($\beta=1$) or negative values only $\beta=-1$.
Therefore, the
results for $\alpha<1$ and $|\beta|=1$ for AA boundaries are the
same as for AR ($\beta=-1$) or RA ($\beta=1$) boundaries. This effect
is clearly visible for the L\'evy--Smirnoff case depicted in Fig.~\ref{smirnoff}.
Now, the results for AA and RA boundaries are the same, as expected. A similar behavior is
observed in Fig.~\ref{mfpt_b10} above, where the values of the MFPTs
for $\alpha<1$ for AA and AR boundaries agree.

\subsubsection*{Survival probability: comparison with Sparre-Andersen scaling}

\begin{figure}[!ht]
\includegraphics[angle=0, width=8.5cm]{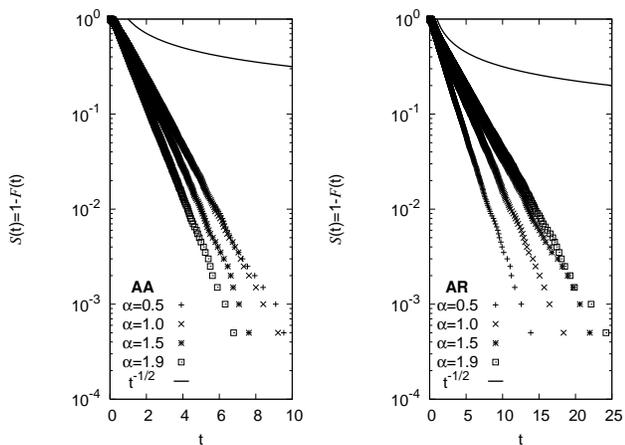}
\caption{Survival probability $S(t)=1-\mathcal{F}(t)$ for confined, symmetric L\'evy--Brownian motion $\beta=0$
 on the interval $B_1=0,B_2=4$ with absorbing half-line for $x<0$ and absorbing (or reflecting) half-line
 for $x > 0$
for various values of the stability index $\alpha$ and midpoint initial
conditions at $x_0= 2$. The left panel depicts the results for AA boundary setups;
the AR cases are displayed with the right panel. The solid line in both data sets
represents the power law $t^{-1/2}$ which describes an asymptotic behavior
foreseen by the Sparre-Andersen theorem \cite{SPARRE} for diffusion on semi-infinite intervals.}
\label{interval}
\end{figure}

The survival probability $\mathcal{S}(t)=1-\mathcal{F}(t)$ for symmetric L\'evy--Brownian motion
on the finite interval
is displayed in Fig.~\ref{interval}.
The motion on a confined support leads to an exponential decay of the survival
probability with the steepness of the slope depending on the stability index $\alpha$.
Notable differences can be observed for various boundary setups:
For motion occurring between two absorbing boundaries, the survival probability
decays faster than for the case when one of the boundaries is taken as reflecting.
It is also useful to emphasize that deviations from the exponential behavior of the survival probability
can be observed in systems subjected to both,
dichotomic and L\'evy stable noises \cite{dybiec2004b,dybiec2004c}, respectively.

\begin{figure}[!ht]
\includegraphics[angle=0, width=8.5cm]{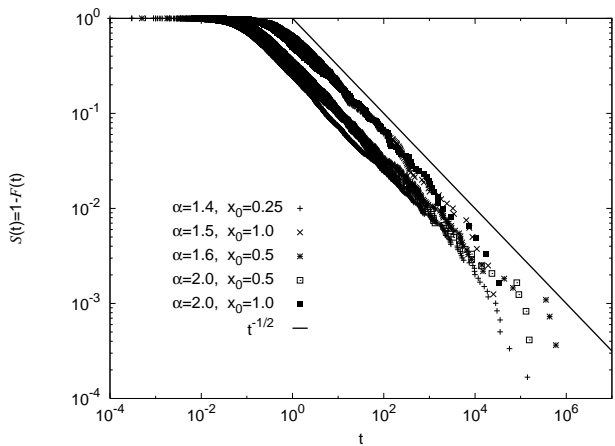}
\caption{Survival probability $S(t)=1-\mathcal{F}(t)$ for free
L\'evy flights on the half line with absorbing boundary for $x<0$
for various values of stability index $\alpha$ and various initial
conditions $x_0$. Remaining simulation parameters like in Fig.~\ref{mfpt_b00}.
The survival probability nicely fits $t^{-1/2}$ slope predicted by Sparre-Andersen scaling.}
\label{anderson}
\end{figure}

Finally, for the systems driven by L\'evy stable noises we also tested the
Sparre-Andersen scaling behavior on an infinite half line. According to the Sparre-Andersen
theorem \cite{SPARRE} for a free stochastic processes driven by symmetric white noises,
the first passage time densities $\frac{d\mathcal{F}}{dt}$,
process from the real half line asymptotically behave like $t^{-3/2}$.
Consequently the survival probability, i.e. the probability of finding
a particle starting its motion at $x_0>0$ in the real half line, scales
like $t^{-1/2}$. In Fig.~\ref{anderson} the survival probability
$S(t)=1-\mathcal{F}(t)$ is depicted for various stability indices $\alpha$ and various
initial conditions $x_0$. It
is clearly visible that the survival probability $S(t)$ behaves like a power law with the
exponent ($ -1/2$), as predicted by the Sparre-Andersen theorem. For the testing
of the Sparre-Andersen theorem the whole negative half line was assumed
to be absorbing.

\section{Conclusions}

With this work we investigated the problem of the mean first passage time and the first passage time statistics
for Markovian, L\'evy--Brownian motion proceeding on a finite interval. The intricate problem of setting
up the proper boundary conditions for absorption and reflection are discussed
with possible pitfalls being pointed out. In particular, it has been demonstrated by numerical studies
that the use of the commonly known, local boundary condition of vanishing flux (in case of reflection) and
vanishing probability (in case of absorbtion), valid for normal Brownian motion
(i.e. $\alpha=2$) no longer apply
for L\'evy white noise. This is so, because the large, jump like excursions of L\'evy
flight increments causes non-continuous, i.e. discontinuous sample trajectories. This in turn requires
the use of nonlocal boundary conditions. It is presently not known how these nonlocal
boundary conditions can be recast as in {\it equivalent} form as a modified,
locally defined differential condition involving the statistical quantities
of interest (the MFPT and the first passage time densities) at the location of
the boundary alone \cite{VDB,weiss83}.

For symmetric, $\alpha$ stable L\'evy white noise driven Brownian motion
we find a bell-shaped, non-monotonic behavior of the MFPTs for absorbing-absorbing
boundaries, with the maximum being assumed for $\alpha=1$. In contrast, for reflecting-absorbing
boundaries we find a monotonic increase. With asymmetric L\'evy white noise, i.e. with a non-vanishing
skewness parameter $\beta \neq 0$ the MFPT results in an even more complex behavior.
As a function of the stability index $\alpha$ the MFPT can exhibit a discontinuous behavior, see the
right panel in Fig.~\ref{mfpt_b10}.

In addition we have studies also the statistics of the first passage times in terms of the
cumulative first passage time distribution function (CDF) and the corresponding
survival probability. In this context, we also
belabor the role of a finite support and different boundary setups for
L\'evy white noise driven Brownian motion for the
universal scaling law of Sparre-Andersen. While the restricted L\'evy--Brownian motion exhibits an
exponential behavior on finite intervals, the crossover to the universal $t^{-1/2}$- power law, being valid
on the half-line for all stability indexes is assumed for large intervals only.


\appendix
\section{$\alpha$-stable random variables}
The random variables $\zeta$ corresponding to the characteristic
functions (\ref{charakt}) and (\ref{charakt1}) can be generated using
the Janicki--Weron algorithm \cite{weron1995,weron1996}. For $\alpha\neq1$, their representation reads
\begin{eqnarray}
\zeta & = & D_{\alpha,\beta,\sigma} \frac{\sin(\alpha(V+C_{\alpha,\beta})) }{
(\cos(V))^{\frac{1}{\alpha}}} \nonumber \\
& \times &
\left[ \frac{\cos(V-\alpha(V+C_{\alpha,\beta}))}{W}
\right]^{\frac{1-\alpha}{\alpha}} \;,
\label{recipe1}
\end{eqnarray}
with the constants $C,D$ given by
\begin{equation}
C_{\alpha,\beta}=\frac{\arctan\left(\beta\tan(\frac{\pi\alpha}{2})\right)}{
\alpha},
\end{equation}
\begin{equation}
D_{\alpha,\beta,\sigma}=\sigma\left[ \cos\left(
\arctan\left(\beta\tan(\frac{\pi\alpha}{2})\right) \right) \right]^{-\frac{1}{\alpha}}.
\end{equation}
For $\alpha=1$, the random variable $\zeta$ can be calculated from the formula
\begin{eqnarray}
\zeta & = & \frac{2\sigma}{\pi} \left[ (\frac{\pi}{2}+\beta V)\tan(V) -\beta\ln
\left( \frac{\frac{\pi}{2}W\cos(V)}{\frac{\pi}{2}+\beta V}
\right) \right].
\label{recipe2}
\end{eqnarray}
In the above equations $V$ and $W$ denote independent random variables; namely,
 $V$ is uniformly distributed in the interval
$(-\frac{\pi}{2},\frac{\pi}{2})$ while $W$ is exponentially
distributed with a unit mean \cite{weron1995}. The numerical
integration scheme has been performed for $\mu=0$ with the increments
of $\Delta x$ (see Eqs.~(\ref{lang}) and~(\ref{lcalka})) sampled from the strictly stable distributions \cite{janicki1994}.

The analytical expressions for stable probability distributions
$L_{\alpha,\beta}(\zeta;\sigma,\mu)$ are known in few cases only:
For $\alpha=0.5,\;\beta=1$ the resulting distribution is the L\'evy--Smirnoff one; i.e.,
\begin{eqnarray}
L_{1/2,1}(\zeta;\sigma,\mu) & = & \left( \frac{\sigma}{2\pi}
\right)^{\frac{1}{2}}(\zeta-\mu)^{-\frac{3}{2}} \nonumber \\
 & \times & \exp\left(-\frac{\sigma}{2(\zeta-\mu)} \right).
\label{smirnoffeq}
\end{eqnarray}
In contrast, for $\alpha=1,\;\beta=0$ one obtains the Cauchy distribution
\begin{equation}
L_{1,0}(\zeta;\sigma,\mu)=\frac{\sigma}{\pi}\frac{1}{(\zeta-\mu)^2+\sigma^2}\;.
\label{cauchy}
\end{equation}
The familiar case with $\alpha=2$ with arbitrary $\beta$ yields the Gaussian PDF.
The prominent characteristic feature of the distributions $L_{\alpha,\beta}(\zeta;\sigma,\mu)$
is its existence of moments up to the order $\alpha$, i.e. the integral
 $\int_{-\infty}^{\infty}L_{\alpha,\beta}(\zeta;\sigma,\mu)\zeta^\alpha d\zeta$
 is finite. This statement results in the conclusion that the only {\it stable}
 distribution possessing a finite second moment is the Gaussian; for all
 other values of $\alpha$ the variance of a stable distribution diverges,
 and for $\alpha < 1$ even the first moment does not exist.

\begin{acknowledgments}
The Authors acknowledge the financial support from the Polish
State Committee for Scientific Research (KBN) through
the grants 1P03B06626 (2004--2005) and 2P03B08225 (2003--2006)
and the ESF funds (E.G.N. and P.H.)
via the STOCHDYN program.
Additionally, BD acknowledges the financial support from
the Foundation for Polish Science through the domestic grant for young scientists (2005).
Computer simulations have been performed at the Academic Computer Center CYFRONET AGH, Krak\'ow.
\end{acknowledgments}

\end{document}